\documentclass[11pt]{article}
\usepackage{graphicx}
\usepackage{amsmath}
\input psfig.sty

\def\kkonium{\textsc{kkonium}}

\begin{document}

\begin{titlepage}

\title{\bf \large THE INTERPLAY OF ULTRAHIGH-ENERGY COSMIC RAYS \\ AND EXTRA
DIMENSIONS \vspace{0.2cm}}

\author{\large Je-An \ Gu\thanks{%
E-mail address: jagu@phys.ntu.edu.tw} \\
{\small Department of Physics, National Taiwan University, Taipei
106, Taiwan, R.O.C.}
\medskip
}

\date{\small \today}

\maketitle

\begin{abstract}
Regarding ultrahigh-energy cosmic rays (UHECRs) as a probe and
extra dimensions as a possible ingredient of the fantastic
ultrahigh-energy world, we discuss possible interplay between
them. On the one hand small extra dimensions and Kaluza-Klein
bursts present a feasible way to the origin of UHECRs. On the
other hand large extra dimensions may change various interactions
between particles at ultrahigh energies and therefore impact on
air showers created by UHECRs. Conversely, UHECR data may tell us
secrets veiled in extra dimensions.
\end{abstract}


\thispagestyle{empty}

\end{titlepage}

\section{Introduction}
The detection of ultrahigh-energy cosmic rays (UHECRs) of energies
above $10^{19}\,$eV, especially super-GZK
(Greisen-Zatsepin-Kuzmin) events, entails various unsolved puzzles
and accordingly brings us serious challenges of understanding
UHECRs: their origins, compositions, unusual largeness of energy,
distribution of arrival directions and times, interactions with
background particles or fields along the journal to the earth,
interactions with the atmosphere and showers generated thereof,
and detection methods. UHECRs are more mysterious under the GZK
mechanism \cite{GZK}, which may lead to the GZK cutoff in the
spectrum, as to be explained later. There is a controversy about
the GZK cutoff: AGASA data show no cutoff \cite{Takeda:1998ps},
while HiRes data admit a cutoff in spectrum around $10^{20}\,$eV
\cite{Abu-Zayyad:2002ta}.  (For a review, see
\cite{Anchordoqui:2002hs}.)

The GZK mechanism is an energy degrading process in the journey of
ultrahigh-energy particles through the universe, caused by the
interactions with background particles or fields such as
background photons $\gamma$ [cosmic microwave background (CMB),
radio background (RB) and infrared background (IRB)], relic
neutrinos $\nu_{\textsc{\scriptsize c$\nu$b}}$, or cosmic magnetic
fields B. In the following we list the attenuation lengths of
various particles of energy $10^{20}\,$eV: (in unit of Mpc)
\begin{center}
\begin{tabular}{cccccc} \hline 
p + $\gamma_{\textsc{\scriptsize cmb}}$ & %
N + $\gamma_{\textsc{\scriptsize cmb,irb}}$ & %
$\gamma$ + $\gamma_{\textsc{\scriptsize cmb,rb}}$ & %
e + $\gamma_{\textsc{\scriptsize cmb,rb}}$ & %
e + B & %
$\nu$ + $\nu_{\textsc{\scriptsize c$\nu$b}}$ \\
\hline %
100 Mpc & %
10 Mpc & %
10--100 Mpc & %
6--13 Mpc & %
2--100 Mpc & %
$\gg 100$ Mpc \\
\hline 
\end{tabular}
\end{center} 
where p, N, e, and $\nu$ stand for proton, nuclei, electron, and
neutrino, respectively.

\vspace{2em}

\section{Theoretical Challenges from Ultrahigh-Energy Cosmic Rays}
Ultrahigh energies and isotropic arrival directions in the
observational results, together with the GZK mechanism, pose the
main challenges for UHECR models, which are usually divided into
bottom-up and top-down models. In the following we discuss the
difficulties of explaining UHECRs in these two categories of
models.

\subsection{Bottom-up Models}
In bottom-up models (for a review, see
\cite{Bhattacharjee:1998qc,Olinto:2000sa}), particles are
accelerated to ultrahigh energies within extreme astrophysical
environments, such as cluster shocks, active galactic nuclei,
neutron stars, and maybe some environment associated with
gamma-ray bursts. Usually these extreme environments are very
dense. How ultrahigh-energy particles escape from these dense
regions without losing much energy through the scattering with
particles therein is a serious intrinsic problem.

In addition, most of particles under the GZK mechanism cannot
maintain energies beyond the GZK threshold (around $10^{20}\,$eV)
after travelling a distance longer than about 50 Mpc, so that it
is unlikely for UHECR sources to be located outside the GZK zone,
a region with a radius of about 50 Mpc around the earth.
Unfortunately, there are very few powerful enough sources within
the GZK zone. These few sources can hardly explain UHECR data, in
particular, the spectrum and the distribution of arrival
directions.

\subsection{Top-Down Models}
In top-down models (for a review, see
\cite{Bhattacharjee:1998qc}), UHECRs are produced via decays of
very massive particles (or topological defects) that may be relics
of the early universe. Therefore energy is not an issue as long as
the mass is large enough. These very massive relic particles might
behave like dark matter and reside in the local dark halo, waiting
for decays that generate UHECRs reaching us. In this case, the
roughly isotropic distribution of arrival directions is not an
issue, either. Nevertheless, these very massive particles are
exotic, i.e., they are beyond the standard model of particle
physics. Their unconfirmed existence is an essential problem in
this kind of models.

In addition, many top-down models involve QCD fragmentation in the
production of UHECRs. UHECRs originating from such fragmentation
are mostly photons and neutrinos, which seem to be disfavored by
present data. However, whether photons or neutrinos can be UHECR
primaries is still not concluded.

\vspace{2em}

\section{Interplay of UHECRs and Extra Dimensions}
So long as UHECR observations have opened the door to
understanding the unknown ultrahigh-energy world, we consider a
possible essential ingredient in this fantastic world, extra
(spatial) dimensions, whose existence is required by various
theories beyond the standard model, especially in the theories for
unifying gravity and other forces, such as superstring theory.


\subsection{Small Extra Dimension: Kaluza-Klein Burst}
Recently a new mechanism for generating UHECRs, by employing small
extra dimensions as a perfect bearer of large energies, is
proposed \cite{Gu:2003wm}. In this model UHECRs are generated via
Kaluza-Klein (KK) bursts, a violent energy transfer from extra
dimensions to ordinary dimensions through collisions between KK
modes.

With the benefit of various features of small extra dimensions,
many difficulties in UHECR models can be overcome, as outlined in
the following. The violent energy transfer from extremely small
extra dimensions to ordinary dimensions through KK bursts can
easily reach required ultrahigh energies, thereby making it
possible to construct top-down models without introducing new
particle species. Against the energy pillage by the GZK mechanism,
KK momentum conservation can protect the energy stored in extra
dimensions when particles travel through the universe. This
feature can help particles of large energies escape from the
source without losing much energy and make it possible for UHECR
sources (more exactly, sources of KK modes which generate UHECRs)
to be located outside the GZK zone, thereby benefitting bottom-up
models.

As shown in \cite{Gu:2003wm}, clumped KK modes of sizes ranging
from KK stellar compact objects to ``\kkonium'', a bound state
consisting of KK modes, are possible origins of UHECRs. The energy
density of these KK stellar compact objects should range from that
of a white dwarf to a neutron star, and the size of \kkonium\
should be around $10^{-6}\,$cm or less, corresponding to the
required size of extra dimensions, $10^{-27}$--$10^{-25}\,$cm.
Small-scale clustering in arrival directions and dispersion in
arrival times of these clustering UHECR events are possible
signatures of KK compact objects.

\subsection{Large Extra Dimension}
In the scenario of large extra dimensions, the cross section of
interactions between various particles may be significantly
changed due to the presence of KK towers of gravitons that mediate
interactions. For example, ultrahigh-energy cosmic neutrinos when
entering the atmosphere may produce black holes, increasing the
neutrino-nucleon cross section and initiating air showers with a
higher rate, thereby making neutrinos more qualified to be the
UHECR primaries
\cite{Feng:2001ib,Anchordoqui:2001ei,Emparan:2001kf}.

However, it is not good for bottom-up models. The same feature
will make ultrahigh-energy particles even more difficult to escape
from their sources without losing much energy, and therefore make
bottom-up models more unlikely to work.

As a remark, the small and the large extra dimensions are not two
contradictory aspects. Our world may possess both. In addition, in
the scenario of large extra dimensions the small thickness of the
brane can play the role of extremely small extra dimensions
employed in \cite{Gu:2003wm}.

\vspace{2em}

\section{Conclusion: an Opening of the Ultrahigh-Energy World}
The detection of UHECRs have opened a window for the unknown
ultrahigh-energy world. Regarding extra dimensions as a possible
substantial ingredient in this world, there may be essential
interplay between them. As suggested in \cite{Gu:2003wm}, small
extra dimensions and KK bursts present a feasible way to
understanding the origin of UHECRs from both bottom-up and
top-down viewpoints. In addition, large extra dimensions may
modify the strength of various interactions and impact on air
showers created by UHECRs. Conversely, the details of UHECRs may
also provide important information (or constraints) about extra
dimensions. Apparently, a great feast in the opening of this
fantastic ultrahigh-energy world is coming soon.

\newpage

\section*{Acknowledgments}
This work is supported by the CosPA project of the Ministry of
Education (MoE 89-N-FA01-1-4).

\end{document}